# Broadband gain characterization of $Co:MgF_2$ for mid-infrared femtosecond pulse amplification

**NATSUKI KANDA,[1,2] KAITO NISHIMIYA,[1] EIJI J. TAKAHASHI[1,*]**

[1] *Extreme Photonics Research Team, RIKEN Center for Advanced Photonics, RIKEN, 2-1 Hirosawa, Wako, Saitama 351-0198, Japan*
[2] *n-kanda@riken.jp*
**Corresponding author: ejtak@riken.jp*

**The broadband gain characteristics of $Co:MgF_2$ were investigated to assess its potential as a gain medium for ultrashort-pulse amplification around the 2 µm spectral region. Single-pass gain measurements performed using femtosecond seed pulses revealed broadband amplification in the 1.5–2.4 µm region. The temporal dynamics and spectra of the gain were experimentally characterized and utilized for numerical simulations to assess the feasibility of $Co:MgF_2$ as a broadband gain medium for ultrashort pulses. The results indicate its potential for broadband amplification in future short-wave infrared to mid-infrared ultrafast laser systems, particularly when combined with coherent waveform synthesis or post-compression techniques.**

## 1. Introduction

Ultrashort laser pulses in the mid-infrared (MIR) spectral region have attracted considerable attention because of their importance in strong-field and ultrafast science, including high-harmonic generation (HHG), ultrafast spectroscopy, and precision laser processing [1]. In particular, few-cycle or near-single-cycle MIR optical pulses are essential for generating isolated attosecond pulses via HHG at higher photon energies because the high-harmonic cutoff energy scales with the square of the driving wavelength [2-5]. Therefore, the development of broadband and high-power ultrashort pulse sources in this spectral region is important for next-generation attosecond science.

Several approaches have been explored to generate ultrashort MIR pulses. Optical parametric amplification (OPA) is a widely used technique that achieves broad-wavelength tunability through nonlinear frequency conversion [6,7]. OPA-based systems have performed remarkably in terms of pulse duration and wavelength coverage.

In contrast, direct laser oscillation and amplification using broadband laser gain media offer important advantages, including potentially higher efficiency, simpler system architectures, improved stability, and direct power scalability. Figure 1(a) shows the gain bandwidths of representative laser materials around the 2 µm spectral region. Among them, Cr:ZnSe has emerged as a promising candidate owing to its extremely broad emission bandwidth and femtosecond pulse generation. Few-cycle and millijoule-class chirped-pulse amplification using Cr:ZnSe has already been demonstrated [8]. In addition to Cr:ZnSe, as shown in Fig. 1(a), $Co:MgF_2$ is particularly attractive because its gain spectrum extends toward shorter wavelengths than Cr:ZnSe, providing complementary spectral coverage for broadband MIR waveform synthesis. In this study, we focused on $Co:MgF_2$ as another promising candidate, whose gain bandwidth covered approximately 1.5–2.5 µm. Although a single gain material cannot support octave-spanning amplification, hybrid approaches combining different broadband gain materials, such as Cr:ZnSe and $Co:MgF_2$, may facilitate high-energy MIR single-cycle optical pulse generation through coherent waveform synthesis [9,10]. To assess this possibility, the gain characteristics of $Co:MgF_2$ must be thoroughly understood.

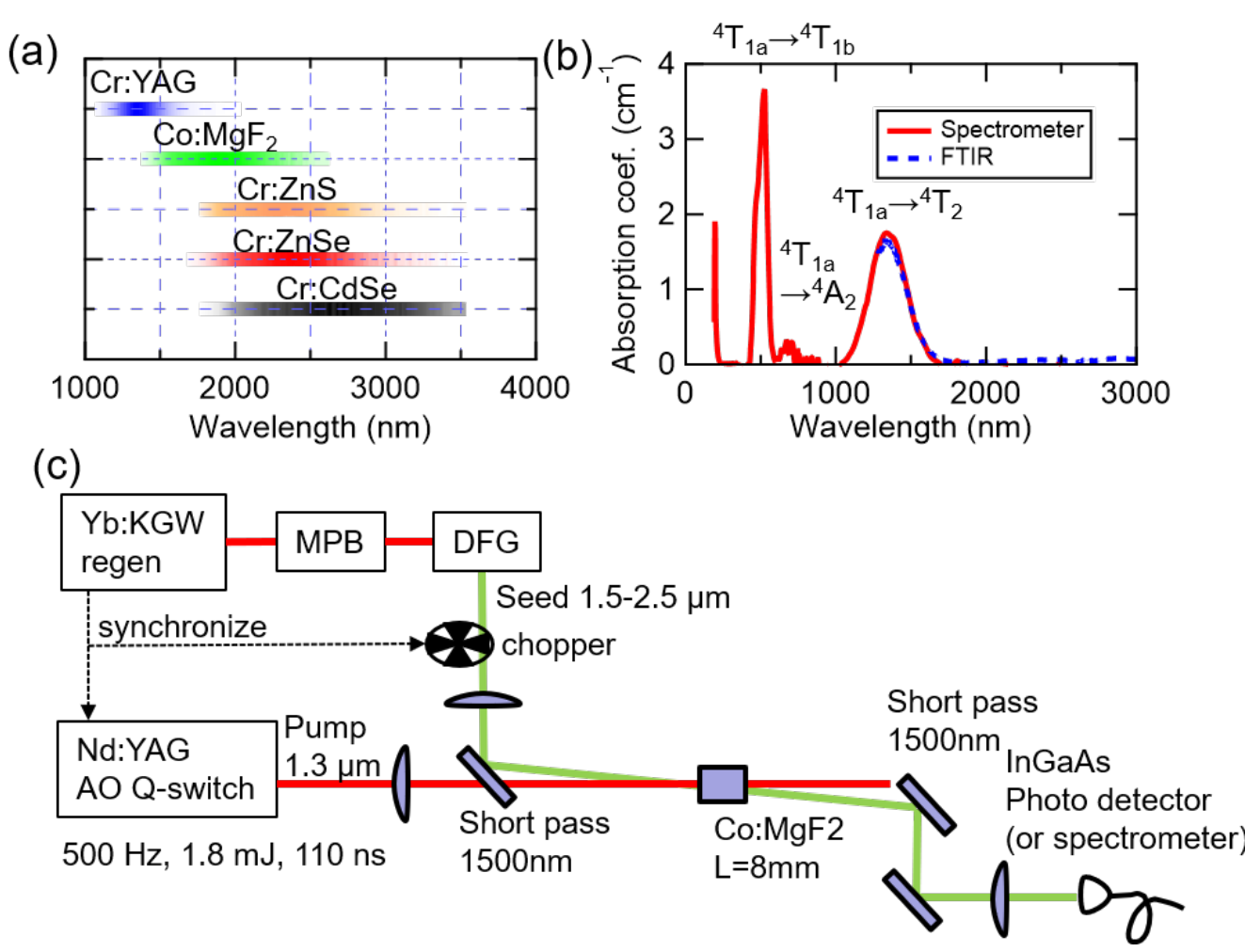

Fig. 1. (a) Comparison of the emission bandwidth for gain materials in the SWIR to MIR regions. (b) Absorption coefficient of 4 wt.% doped $Co:MgF_2$ crystal with thickness of 2 mm, as measured using a spectrometer and FTIR. (c) Schematic of the setup for measuring single-pass gain in $Co:MgF_2$ crystal. MPB: multi-plate broadening, DFG: difference frequency generation.

Historically, $Co:MgF_2$ has attracted considerable interest as a broadband transition-metal-doped laser material. The exceptionally broad emission spectrum of $Co:MgF_2$ originates from strong vibronic broadening and the splitting of the ground state into six sublevels, resulting in a remarkably wide gain bandwidth [11]. This characteristic was investigated extensively in the pioneering work of P. F. Moulton. Furthermore, $Co:MgF_2$ exhibits absorption around the 1.3 μm wavelength region, which enables pumping with mature and reliable Nd:YAG-based laser sources and provides an additional practical advantage.

Despite the relatively small emission cross-section (~$10^{-21}$ $cm^2$) of $Co:MgF_2$, its favorable thermomechanical properties, including relatively high thermal conductivity, high Mohs hardness, and small thermo-optic coefficients [12, 13], have enabled watt-level average-power operation. Owing to its broad gain bandwidth, $Co:MgF_2$ has been investigated as a widely tunable laser source in both continuous-wave [14-16] and pulsed regimes [13,17,18]. Additionally, actively mode-locked picosecond laser oscillations have been reported [19].

These characteristics suggest that $Co:MgF_2$ is a promising candidate for broadband ultrashort pulse amplification and power scaling in the MIR region. Nevertheless, to the best of our knowledge, no studies have demonstrated the amplification of femtosecond laser pulses using $Co:MgF_2$. As a first step toward hybrid broadband amplification and coherent waveform synthesis in the 1.5–3 μm spectral region, we investigated the broadband gain characteristics of $Co:MgF_2$ using femtosecond seed pulses.

## 2. Experiments and Results

To evaluate the gain properties, a 4 wt.% Co-doped $MgF_2$ crystal with a thickness of 8 mm was used in this study. The surfaces were optically polished and uncoated. This material has an absorption peak around 1.3 μm as shown in Fig. 1(b). A homebuilt Q-switched Nd:YAG laser operating at dual wavelengths of 1319 nm and 1338 nm was developed as the pump source [20,21]. Pulses with a duration of 110 ns, pulse energy of 1.8 mJ at a repetition rate of 1 kHz, and beam quality $M^2$ of (1.07, 1.03) in the (x, y) directions were obtained. The timing of the optical pulses was tuned by controlling the timing of the acousto-optic modulator and quasi-CW pump laser diode modulation which were synchronized with the external trigger signals.

Intra-pulse different frequency generation (DFG) in a $BiB_3O_6$ crystal was used for the broadband MIR seed pulse. For the fundamental light source, sub-10 fs pulses were obtained via two-stage multi-plate broadening [22-24] using an Yb:KGW regenerative amplifier (PHAROS, Light Conversion, UAB).

Figure 1(c) shows the setup for evaluating the single-pass gain. Both pump and seed pulses were focused on the $Co:MgF_2$ sample at a small collinear angle (approximately 2°). The $1/e^2$ beam diameter of the pump was 143 μm and

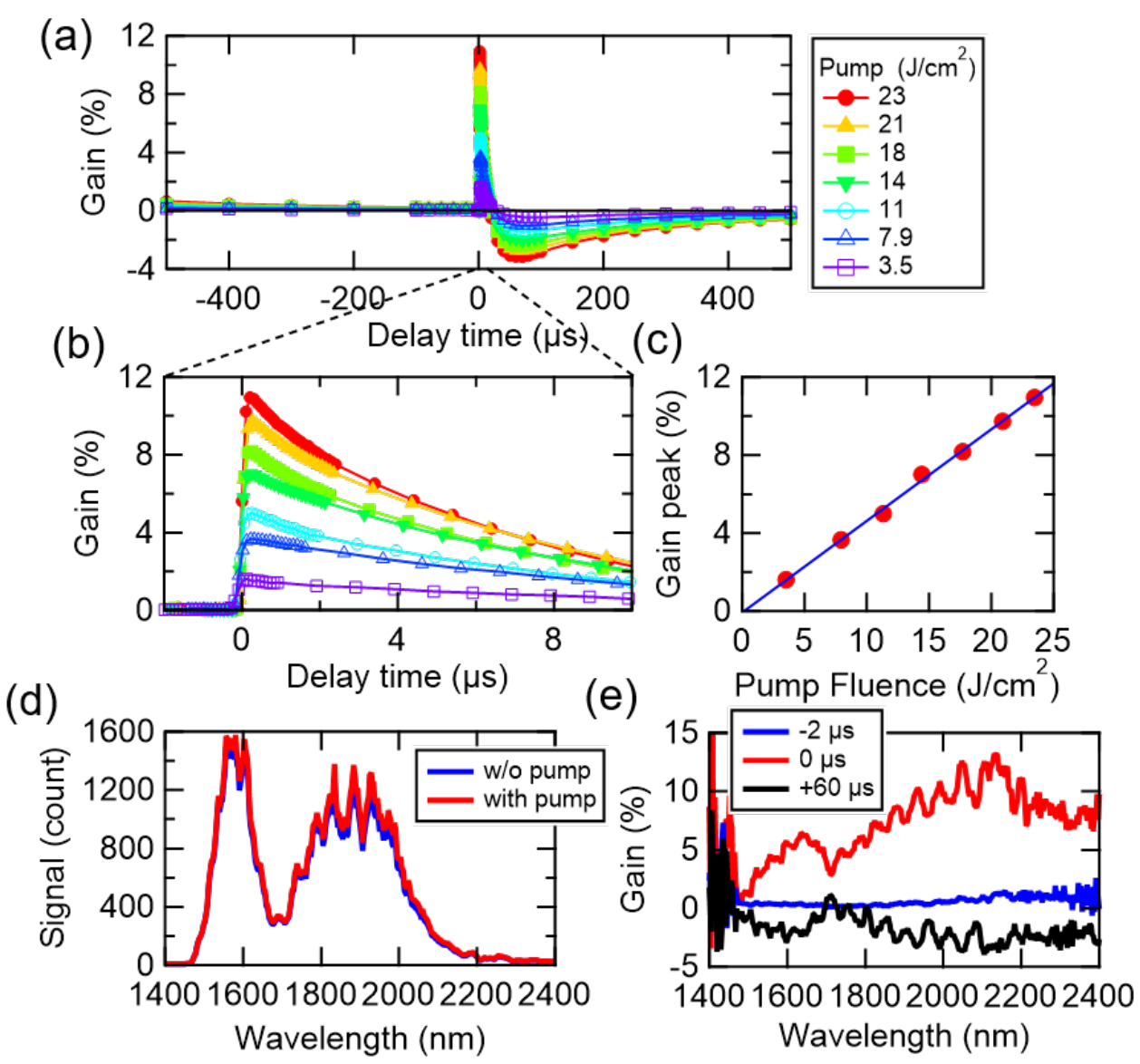

Fig. 2. Experimental results of single-pass gain measurement. (a) Dynamics of the gain signals for different pump fluences, (b) expansion of the data of (a), in the range from -2 to +10 μs. (c) Pump fluence dependence of single-pass gain signal, with the red dots and blue line representing the measured values and linear fit, respectively. (d) Measured spectra for the probe pulses with and without the pump pulse at zero-delay time. (e) Single-pass gain spectra for different pump-probe delay times.

133 μm in the x and y directions, respectively, and the beam size of the probe was slightly smaller.

The pump and seed were modulated with different frequencies to extract the gain signal with high precision. The Yb:KGW laser was operated at 1 kHz, and the seed DFG was modulated with 250 Hz using an optical chopper. The 1.3 μm pump laser was operated at 500 Hz, which was half the DFG with synchronization. The transmitted seed pulses were detected shot-by-shot using an InGaAs photodiode or spectrometer. This pulse-resolved measurement was used to separate the dataset into 4-types: pump on/off and seed on/off. The systematic background signals were eliminated by subtraction.

Figure 2(a) and (b) show the dynamics of the single-pass gain signal measured using a photodiode. The gain signals exhibit a maximum at the zero pump-seed delay between the pump and seed, and decay after the pump. The decay time is around 14 μs, which is comparable with the lifetime of this material, 36 μs at room temperature [11]. After this decay, the gain signal overshoots to negative, which indicates loss, with longer recovery time of approximately 270 μs. However, the origin of this long-lived loss remains unclear. Possible mechanisms include excited-state absorption and thermally induced changes in the crystal. This long-lived loss component may limit efficient operation at higher repetition rates.

Figure 2(c) shows the pump fluence dependence of the peak gain values. A gain of 12% is observed at a pump fluence of 23 J/cm$^2$, which is almost proportional to the pump fluence in this region. To the best of our knowledge, this is the first demonstration of broadband femtosecond pulse amplification in $Co:MgF_2$.

To evaluate the gain spectrum, the pulse-resolved measurements were performed using a spectrometer (NIR-QUEST; Ocean Optics) at a pump fluence of 23 J/cm$^2$. As shown in Fig. 2(d), the spectrum is slightly enhanced by pumping at zero delay time. Figure 2(e) shows the gain spectra for different delay times. At a negative delay time (−2 µs), the gain spectrum is almost zero for all the wavelengths, whereas at zero-delay, broadband gain is observed in the 1.5–2.4 µm region. The peak gain is approximately 13%, which is consistent with the photodiode measurements. Subsequently, at a delay time of 60 µs, a negative spectrum is observed, similar to the photodiode measurements. The observed gain bandwidth supports near-two-cycle pulse durations and demonstrates the potential of $Co:MgF_2$ as a broadband gain medium for MIR waveform synthesis.

### 3. Discussions

We estimated the expected gain based on the reported emission cross section and compared it with the experimental results. The peak emission cross section in this wavelength region is $\sigma_{em} = 1.0 \times 10^{-21}$ cm$^2$ [6]. In our experiments, the crystal length $L$ was 8 mm, and the absorption of the pump at this length was 70%, as measured using a spectrometer (V570, JASCO Corp.). Assuming a quantum efficiency of unity, the excited ion density is estimated as $N = 1.2 \times 10^{20}$ cm$^{-3}$. Neglecting reabsorption, the one pass transmittance is expressed as $\exp(\sigma_{\rm em} NL) = \exp(0.096) = 1.10$, which is 10% of the single-pass gain. This value was consistent with our experimental measurements. Although the single-pass gain should be exponentially dependent on the excitation density $N$, the experimental dependence exhibited proportional behavior (Fig. 2(c)) because $\sigma_{\rm em} NL \ll 1$ and the approximation $1 - \exp(\sigma_{\rm em} NL) \approx \sigma_{\rm em} NL$ was applicable.

Finally, the possibility of its application in the master oscillator power amplifier is discussed. As the measured single-pass gain was relatively modest, a large number of passes would be required in conventional multipass amplifier configurations. Therefore, the possibility of using regenerative amplifiers (RA) was considered. We numerically simulated the amplification in the RA based on the Frantz-Nodvik equation [25]. The stored and laser fluences of each round trip were expressed using the following recurrence formula:

$$J_{sto}^{(n+1)} = J_{sto}^{(n)} - \left(J_{out}^{(n)} - J_{in}^{(n)}\right). \quad (1)$$

Here, $J_{sto}^{(n)}$ is the stored fluence, and $J_{in}^{(n)}$ and $J_{out}^{(n)}$ are the laser fluences before and after the gain medium in the $n$-th round trip, respectively. Considering the round-trip throughput factor $\Gamma$, the input to the gain medium in next round-trip can be expressed as $J_{in}^{(n+1)} = \Gamma J_{out}^{(n)}$. Considering the saturation

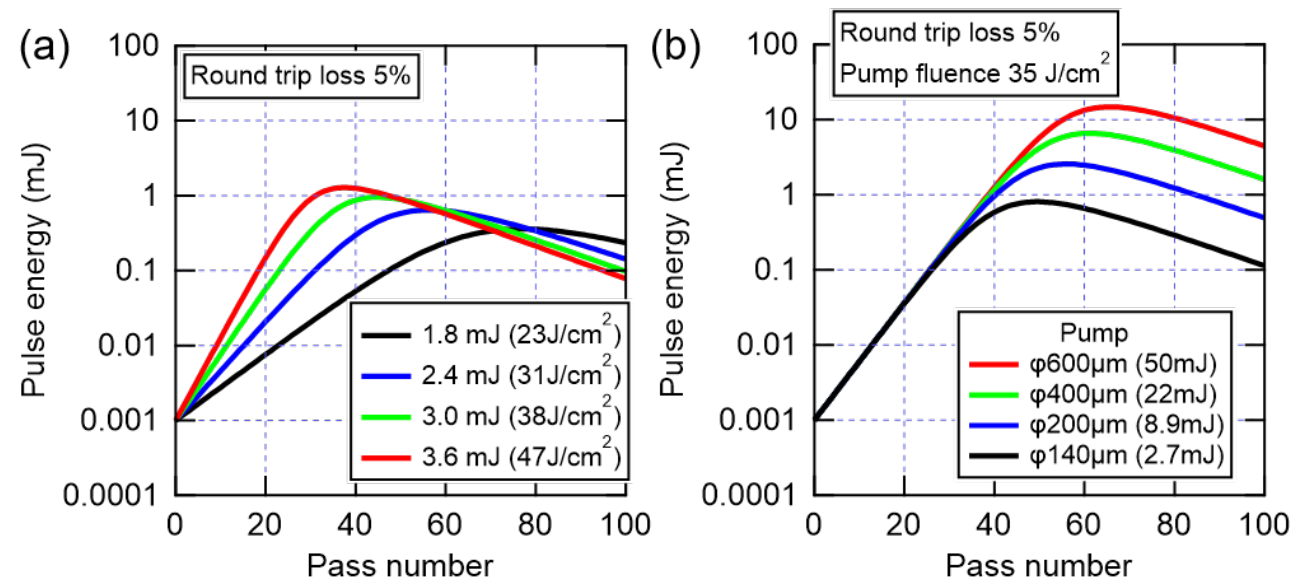

Fig. 3. Numerical simulation of amplification based on the Frantz-Nodvik equation. (a) Calculation results for a fixed beam diameter of 140 µm. (b) Calculation results for beam size scaling at a fixed pump fluence of 35 J/cm$^2$.

effect, the amplification of each step is expressed as follows:

$$J_{out}^{(n+1)} = J_{sat} \ln\left[1 + \exp\left(\frac{J_{sto}^{(n)}}{J_{sat}}\right)\left\{\exp\left(\frac{J_{in}^{(n+1)}}{J_{sat}}\right) - 1\right\}\right], \quad (2)$$

where $J_{sat} = \hbar\omega_0/\sigma_{em}$ is the saturation fluence, expressed in terms of the emission angular frequency $\omega_0$ and emission cross section $\sigma_{em}$. The initial condition of stored energy is expressed as $J_{sto}^{(0)} = \eta \frac{\lambda_{pump}}{\lambda_0} J_{pump}$, where $J_{pump}$ is the pump fluence; $\eta$ is the quantum efficiency; and $\lambda_{pump}$ and $\lambda_0$ are the pump and emission wavelengths, respectively. Here, we assumed $\eta = 1$. In the case of $\lambda_0 = 2$ µm and $\sigma_{em} = 1.0 \times 10^{-21}$ cm$^2$, the saturation fluence was 99 J/cm$^2$. Figure 3(a) shows the calculation results for different pump pulse energies. In this simulation, the pump diameter, initial pulse energy of the seed, and round-trip loss were 140 µm, 1 µJ, and 5% ($\Gamma = 0.95$), respectively. In the case of our experimental pump fluence of 23 J/cm$^2$, the pulse energy reaches 0.36 mJ after 77 roundtrips. If the available pump pulse energy is larger, the amplification ratio increases, and the maximum pulse energy increases with smaller round-trip numbers. The calculation indicates that a 1-mJ class output is possible when the pump fluence exceeds 38 J/cm$^2$. However, in our experiments, the $Co:MgF_2$ crystal was accidentally damaged at a pump fluence of 40 J/cm$^2$ when the pulse beam was focused more tightly on the crystal. This pump fluence was slightly lower than the previously discussed damage thresholds of 80 J/cm$^2$ or 400 MW/cm$^2$ [26]. This difference may be attributed to the quality of the crystal or surface conditions. Considering the limitations of the pump fluence, the pump beam size scaling was calculated (Fig. 3(b)) by fixing the pump fluence at 35 J/cm$^2$. The accessible pulse energy increased with the pump pulse energy. In the case of 50 mJ pumping, the maximum amplified pulse energy was 15 mJ for 65 round trips. While our end-pumped Nd:YAG laser possesses a maximum pulse energy of 1.8 mJ, a much higher pulse energy 1.3 µm Nd:YAG Q-switched laser has been reported with a side-pumped configuration, delivering a pulse energy of 119 mJ at a repetition rate of 400 Hz, with pulse duration of 117 ns [27]. Such developments suggest that pump energies on the order of several tens of millijoules may be achievable in future high-power 1.3 µm Nd:YAG laser systems. In addition, achieving broadband electro-optic switching covering the entire 1.5–2.5 μm gain bandwidth remains an important technical challenge.

To access the possibility for the amplification of ultrashort pulses, the inverse Fourier transform of the gain spectrum (Fig. 2(e)) was calculated. The duration of Fourier-transform-limited pulse was 13.5 fs at a center wavelength of 1925 nm, which corresponds to 2.1 optical cycles. The broad gain bandwidth demonstrated here suggests the possibility of employing $Co:MgF_2$ as a complementary amplifier channel to Cr:ZnSe in future coherent waveform synthesis architectures.

## 4. Conclusion

In conclusion, we demonstrated broadband femtosecond pulse amplification in $Co:MgF_2$ for the first time using femtosecond seed pulses around the 2 μm spectral region. The dynamics of the single-pass gain were systematically investigated by controlling the temporal delay between the seed pulses and the Q-switched Nd:YAG pump laser. Spectral measurements confirmed the amplification of femtosecond pulses over a broad spectral range of 1.5–2.4 μm. The measured gain characteristics were consistent with previously reported spectroscopic data obtained using independent methods. Numerical simulations based on the Frantz–Nodvik equation were also performed to evaluate the potential of $Co:MgF_2$ for ultrashort-pulse amplification. Although the relatively small gain, long recovery time, and limited damage threshold of $Co:MgF_2$ may restrict its applicability to high-gain and high-repetition-rate amplification systems, the broad gain bandwidth extending across the MIR region remains an attractive feature. To the best of our knowledge, the present results provide the first assessment of $Co:MgF_2$ for femtosecond pulse amplification and establish a foundation for future studies of broadband MIR gain media. In particular, $Co:MgF_2$ has potential applications as a useful gain medium in hybrid ultrafast laser architectures employing coherent waveform synthesis or external spectral-broadening techniques for the generation of few-cycle optical pulses.

## Funding

This research was supported by the RIKEN TRIP initiative (Leading-edge semiconductor technology), the MEXT Quantum Leap Program (MEXT Q-LEAP) under Grant No. NO118602, JST (K Program) Japan (Grant No. JPMJKP24M1), JST PRESTO (JPMJPR2006), and JSPS KAKENHI (JP24K08293).